\documentclass[prd, nofootinbib, superscriptaddress, tightenlines,
preprintnumbers]{revtex4}
\usepackage{epsfig,latexsym,amssymb,amsfonts}
\usepackage{subfigure}

\begin{document}
\begin{flushright}
  HU-EP-10/31

\end{flushright}
\title{Dynamical Fermion Masses Under the Influence of Kaluza-Klein
Fermions and
a Bulk Abelian Gauge Field}
\author{D. Ebert}
\email{debert@physik.hu-berlin.de}
\affiliation{Institute of Physics, Humboldt-University Berlin,
12489 Berlin, Germany}
\author{V.Ch. Zhukovsky}
\email{zhukovsk@phys.msu.ru}
\author{A.V. Tyukov}
\affiliation{Faculty of Physics, Department of Theoretical Physics,
Moscow State University, 119991, Moscow, Russia}

\begin{abstract}

The dynamical fermion mass generation on 
a 3-brane in the 5D
space-time is discussed in a model with bulk fermions in interaction
with fermions on the 
brane assuming the presence of a constant
abelian gauge field 
component $A_5$ in the bulk. We calculate the effective
potential  as a function of the fermion masses and the gauge field
component $A_5$. The masses can be found from the stationarity condition for the
effective potential (the gap equation). We formulate the
equation for the mass spectrum of the 4D--fermions. The phases with
finite and vanishing fermion masses are studied and the dependence of
the masses on the radius of the 5th dimension is analyzed. The
influence of the $A_5$-component of the
gauge field on the symmetry
breaking is considered both when this field is a background parameter
and a dynamical variable. The critical values of the $A_5$ field,
the coupling constant and the radius are examined.
\end{abstract}

\maketitle
\section{Introduction}

As originally proposed by Kaluza and Klein \cite{KK}, an
extra-dimensional space-time may be compactified to
leave the 4-dimensional space-time as our
real world. It was later demonstrated that the
fundamental mass scale of the compactified space may be much smaller
than the Planck
mass
\cite{Ant,Ark1}. This stimulated intense studies
of phenomenological evidences of extra-dimensional effects
\cite{Han}.Thus in the analysis of \cite{abe}, bulk fermions were
introduced in the 5D space-time interacting with fermions
living
on
a 3-brane.
The interaction between fermions generated as a result of the exchange
of the Kaluza-Klein excited modes of the graviton may be expressed as
effective four-fermion interaction \cite{Han}
that may lead to the dynamical generation of
fermion masses \cite{Dob,Chen}.

On the other hand, there exists also the idea that the Higgs particle
may originate from extra-dimensional components of gauge fields
\cite{gaugehiggs}. In this way, a Yukawa coupling to the 4D
scalar, $A_5$, of the same strength as the gauge coupling,
the so-called gauge-Yukawa unification, may also lead to a mass
generation. A non-zero expectation value of $A_5$
then necessarily breaks the gauge and chiral symmetries of the
underlying theory and $A_5$ thus plays the role of
the Higgs field. This is a sort of radiative symmetry
breaking \cite{Su}, or Hosotani mechanism \cite{hosotani}.
Thus, it is the dynamical symmetry breaking by the four-fermion
interaction which together with this Higgs-type mechanism generate the
physical fermion mass spectrum.

In this paper, we will consider an extension of the fermion model
\cite{abe} describing the interaction of bulk fermions with fermions
on
a 3-brane 
in 5D space-time with one extra dimension
compactified on a circle of radius $R$, by assuming the
additional presence of a constant abelian gauge field  
component
$A_5$  in the
bulk. In particular, we will study
chiral symmetry breaking and
mass generation for light fermions under the
combined influence of the effective four-fermion interactions and the
bulk field 
component
$A_5$ (in the following denoted as gauge field). For this aim, we first derive the effective
potential $V_{\rm eff}(\sigma,A_5)$ as function of the scalar (fermion
condensate) field $\sigma$ and the constant field $A_5$ for the case
of both periodic and antiperiodic boundary conditions of the bulk
fermion field. The mean field values of these fields are then
determined from the corresponding gap equations. On this basis the
critical coupling
constants
$g_c$ for the chiral phase transition and
the fermion condensate are numerically calculated and presented as
functions of the compactification radius $R$ and $A_5$. Finally, we
discuss the resulting fermion mass spectrum obtained from solving the
corresponding eigenvalue equation for ground state and excited
Kaluza-Klein states of fermions.

\section{The model}

Two mechanisms of fermion mass generation are known: the dynamically
generation of fermion masses and the Kaluza-Klein mechanism for
generating masses of excited modes of the bulk fermions due to
compactification. The mass of eventually existing Kaluza-Klein excited
modes is expected to be of the order of TeV \cite{Cha}.
In the following we shall consider
a 5D fermion
model
containing two types of fermion fields
and
a bulk gauge field $A_{M}$. One of the fermions, $\psi$,
lives in the 5-dimensional bulk space, while the other one, $L$,
exists only on a 3-brane which resides at a fixed point of
the extra dimension. The interaction between these fermions is assumed
to be given by a four-fermion term. Let us further suppose
that the bulk fermion $\psi$ is charged and thus may interact with the
bulk abelian gauge field $A_{M}$, whereas the fermion $L$ is
neutral.
Let us consider
the vacuum averages $\langle A_{\mu}\rangle=0$, while $\langle A_5
\rangle=A_5={\rm const}$. 
The presence of the constant potential $A_{5}$
then influences the dynamical mass generation arising from the
four-fermion interaction. 
Moreover, for applying the $1/N_f$ expansion
technique, fermions are assumed to be multiplets of a flavour $O(N_f)$
group and thus to have $N_f$ flavour components.

The model
is described by a straightforward
extension of the Lagrangian studied in \cite{abe}\footnote{We shall
mainly
follow the
notations of this work.}:
\begin{equation}
\label{lagr}
\mathcal{L}^{(5)}=\bar{\psi}i\Gamma^{M}D_{M}\psi +
[\bar{L}i\gamma^{\mu}\partial_{\mu}L - \frac{g^2}{N_f}
(\bar{\psi}\Gamma^{M}L)(\bar{L}\Gamma_{M}\psi)]\delta(y),
\end{equation}
where $M=\mu,\,5;\,\,\mu=0,1,2,3$, $\Gamma^{M}=\{\gamma^{\mu}, i\gamma^5\}$ and
$D_{M}=\partial_{M}-i e A_{M}$ is the
covariant derivative with $A_{M}$ being a bulk abelian gauge field
\footnote{Note that for 5 (odd) dimensions there exists no
  $\gamma_5$-type matrix anticommuting with all other $\gamma$
  matrices like in 4 dimensions and thus no standard chiral
  symmetry. An irreducible representation of a 5-dimensional fermion
  field is given
by a 4-component field as in the case of 4 dimensions, and the fifth
component of the $\gamma$ matrix is just $i\gamma_5$ in 4
dimensions.}. The fifth coordinate $y$ varies in the interval $[0,2\pi
R]$.

Clearly, in the absence of
the gauge field $A_M$ the Lagrangian is invariant
under the
$Z_2$ discrete
``chiral'' symmetry:
$y\to -y$, $\psi(y)\to
\gamma^5\psi(-y)$, $L(y)\to \gamma^5 L(-y)$. The
discrete
symmetry thus prevents a mass term of the form $\bar{\psi}L$. To preserve this symmetry, the field $A_M$ should
transform as $A_{\mu}(y)\to A_{\mu}(-y)$ and $A_{5}(y)\to -A_{5}(-y)$. Therefore the presence of
the constant potential $A_5$ spontaneously breaks the
chiral
symmetry.
In particular it is interesting to
consider the influence of
a constant $A_5$
on the
dynamical mass generation of fermions.

As usually done in the case of NJL-type models (cf. e.g. \cite{EbR}),
we next
introduce an auxiliary
(flavour-singlet)
boson field $\sigma_M$ to obtain a linearized version
of the above model Lagrangian
\begin{equation}
\mathcal{L}_1^{(5)}=\bar{\psi}i\gamma^{\mu}\partial_{\mu}\psi
-\bar{\psi}\gamma^5\partial_5\psi+ i e \bar{\psi}A_5\gamma^5\psi +
[\bar{L}i\gamma^{\mu}\partial_{\mu}L + N_f \sigma^M \sigma_M^{*} + (g
\sigma_M \bar{\psi}\Gamma^{M}L + h.c.) ]\delta(y),
\end{equation}
where summation over Dirac and flavour indices is implicitly included. In the following we shall consider the  dynamical mass generation in the mean field approximation supposing that only the flavor-singlet field $\sigma_5$ acquires a non-vanishing vacuum expectation value $\langle\sigma_5\rangle=-\sigma$,
whereas all the other irrelevant components of $\sigma_M$ vanish,
$\langle\sigma_{\mu}\rangle=0$,
$\mu=0,\ldots,3$.
After performing the chiral rotation $\psi\to e^{i \frac{\pi}{4}\gamma^5}\psi$ and $L\to
e^{i \frac{\pi}{4}\gamma^5}L$
one obtains
\begin{equation}
\mathcal{L}_1^{(5)}=i\bar{\psi}\not\!\partial\psi
-i\bar{\psi}\partial_5\psi - e A_5\bar{\psi}\psi +
[i\bar{L}\not\!\partial L - N_f |\sigma|^2 + (g \sigma \bar{\psi}L +
h.c.) ]\delta(y).
\end{equation}

Since the coordinate $y$ varies on the circle,
the end points $y=0$ and $y=2\pi R$ must be identified. We adopt the general boundary condition for the fifth coordinate
\begin{equation}\label{bc}
    \psi(x,y+2\pi R)=e^{i\pi \delta}\psi(x,y),
\end{equation}
where the phase $\delta$ may have an arbitrary value
in the range $0\leq\delta<2$.
(Note that a change of the phase of the $\psi$-field does
not affect the phase of the
condensate $\sigma$ which exists
only at $y=0$.) In particular, $\delta=0$ corresponds to
a periodic boundary condition
and $\delta=1$ to an antiperiodic one. Then we
obtain the decomposition of the bulk fermion field into the Kaluza-Klein modes
\begin{equation}
\psi(x,y)=N \sum_{n=-\infty}^{\infty} \psi_n(x) e^{i
\frac{y}{R}(n+\frac{\delta}2)},
\end{equation}
where $N$ is a normalization constant.

The Lagrangian (\ref{lagr}) is invariant under the usual
``small''
gauge transformations
\begin{eqnarray}
   A_M(x,y) &\to& A_M(x,y)+\frac1e\partial_M \alpha(x,y), \\
   \psi(x,y) &\to& e^{i\alpha(x,y)}\psi(x,y),
\end{eqnarray}
with periodic function $\alpha$: $\alpha(x,y+2\pi R)=\alpha(x,y)$.

In addition there are also
``large''
gauge transformations which satisfy the condition
\begin{equation}
\alpha(x,y+2\pi R)=\alpha(x,y)+2\pi l, \quad l=\pm1,\pm2,\ldots,
\end{equation}
so that the boundary condition (\ref{bc}) for the transformed field
$\psi$ is not violated.
In particular, if we take $\alpha(x,y)=y l /R$, then
the transformation of gauge fields yields:
\begin{equation}
A_5\to A_5+\frac{l}{eR}, \quad l=\pm1,\pm2,\ldots.
\label{large}
\end{equation}
Thus, an {\it arbitrary} constant potential $A_5$
cannot be gauged away when the fifth coordinate is compactified.

Integration over the coordinate $y$ yields the effective
four-dimensional
Lagrangian
\begin{equation}
\mathcal{L}^{(4)} \equiv \int_0^{2\pi R} dy \mathcal{L}_1^{(5)}=
 \sum_{n=-\infty}^{\infty} \left(i\bar{\psi}_n\not\!\partial\psi_n
+(\frac{n+\delta/2}{R}-a)\bar{\psi}_n\psi_n\right) +
i\bar{L}\not\!\partial L - N_f |\sigma|^2 + \left(m
\sum_{n=-\infty}^{\infty} \bar{\psi}_n L + h.c.\right) ,
\end{equation}
where $a \equiv e A_5$ and $m \equiv N g \sigma$. To obtain the
properly normalized kinetic term, we chose $N=1/\sqrt{2\pi R}$.

\section{The mass spectrum}
Let us next
introduce matrix notations for
the fermion fields
$\Psi^i$ ($i=1,\ldots,N_f$)  \cite{abe}
\begin{equation}
    (\Psi^{i})^{T}=\{L, \psi_0,\psi_1,\psi_{-1},\psi_2,\psi_{-2},\ldots\}^i,
\end{equation}
and for the mass matrix $\mathcal{M}\equiv M\times I_f$, where $I_f$ is the unit operator in flavor space and
\begin{equation}
    M\equiv\left(
             \begin{array}{ccccccc}
               0 & m^{*} & m^{*} & m^{*} & m^{*} & m^{*} & \ldots \\
               m & \frac{\delta}{2R}-a & 0 & 0 & 0 &  0 & \ldots \\
               m & 0 & \frac1R+\frac{\delta}{2R}-a & 0 & 0 & 0 & \ldots \\
               m & 0 & 0 & -\frac1R+\frac{\delta}{2R}-a & 0 & 0 & \ldots \\
               m & 0 & 0 & 0 & \frac2R+\frac{\delta}{2R}-a & 0 & \ldots \\
               m & 0 & 0 & 0 & 0 & -\frac2R+\frac{\delta}{2R}-a & \ldots \\
               \vdots & \vdots & \vdots & \vdots & \vdots & \vdots & \ddots \\
             \end{array}
           \right).
\end{equation}
Then the effective Lagrangian can be written in the
compact matrix form as follows
\begin{equation}
\mathcal{L}^{(4)}= \bar{\Psi}i\not\!\partial\Psi+\bar{\Psi}\mathcal{M}\Psi -
N_f|\sigma|^2.
\end{equation}
It should be noted
that
the fermions represented by  $\Psi$
cannot
be
observed by themselves. Only the eigenstates of the mass matrix $M$ are observable fermions. Supposing that $\sigma_5$ acquires a non-vanishing vacuum expectation value $\langle\sigma_5\rangle=-\sigma$, we expect that the eigenvalues of the matrix $M$ with $m = N g \sigma$ determine the masses of 4-dimensional fermions. They can be found in the following way. Consider the matrix
\begin{eqnarray}
  M - \lambda I\equiv
       \left(
         \matrix{
            -\lambda & m^{*} & m^{*} & m^{*} & m^{*} & m^{*} & \cdots \cr
            m & -\lambda-a +\frac{\delta}{2R}    & 0     & 0     & 0     & 0     & \cdots \cr
            m & 0     & -\lambda-a+\frac{1}{R}+\frac{\delta}{2R}   & 0     & 0     & 0 & \cdots  \cr
            m & 0     & 0     & -\lambda-a - \frac{1}{R}+\frac{\delta}{2R} & 0     & 0 & \cdots  \cr
            m & 0     & 0     & 0     & -\lambda-a+\frac{2}{R}+\frac{\delta}{2R}   & 0 & \cdots  \cr
            m & 0     & 0     & 0     & 0     &-\lambda-a - \frac{2}{R}+\frac{\delta}{2R} &\cdots \cr
            \vdots & \vdots & \vdots & \vdots & \vdots & \vdots & \ddots
         }
       \right). \label{M'}
\end{eqnarray}
Now we make use of the following identity \cite{Madelung}
\begin{eqnarray}
          \left|
\matrix{
            a_{11} & a_{12} &\cdots & a_{1n}& u_1 &  \cr
            a_{21} & a_{22} & \cdots & a_{2n}&u_2 &  \cr
            \vdots & \vdots & \vdots & \vdots & \vdots & \cr
            a_{n1} & a_{n2}   & \cdots & a_{nn}     & u_n & \cr
            v_1 & v_2     & \cdots     & v_n & w &  \cr
       }
       \right| \qquad = \qquad w|a_{ik}|\,-\,\sum_{i=1}^n\sum_{k=1}^n A_{ik}u_iv_k,
\label{Mad}
\end{eqnarray}
where $A_{ik}$ is the algebraic supplement of $a_{ik}$.
As a result, we have
\begin{equation}\label{Eigen'}
\det(M-\lambda I)= \prod_{j=1}^{\infty}\left[
(\lambda+a-\frac{\delta}{2R})^2-\left(\frac{j}{R}\right)^2\right]
\left\{ \lambda(\lambda+a-\frac{\delta}{2R}) -|m|^2 -2
|m|^2(\lambda+a-\frac{\delta}{2R})^2
\sum_{l=1}^{\infty}\frac{1}{(\lambda+a-\frac{\delta}{2R})^2-\left(\frac{l}{R}\right)^2}
\right\}.
\end{equation}
The eigenvalues of the matrix $M$ can
then be obtained from the equation
\begin{equation}
\det ( M - \lambda I )\,=\,0.
\end{equation}
Finally, by using the summation formula
\begin{equation}\label{sum}
    \sum_{n=1}^{\infty}
    \frac1{n^2-c^2}=\frac1{2c^2}-\frac{\pi}{2c}\cot{\pi c},
\end{equation}
one can rewrite the eigenvalue equation as
\begin{equation}\label{det}
\prod_{j=1}^{\infty}\left[
(\lambda+a-\frac{\delta}{2R})^2-\left(\frac{j}{R}\right)^2\right]
\left\{ \lambda(\lambda+a-\frac{\delta}{2R}) - |m|^2[\pi R
(\lambda+a)-\frac{\pi\delta}{2}]\cot[\pi R(\lambda+a)-\frac{\pi
\delta}{2}] \right\}=0.
\end{equation}
It is easy to verify that $\lambda$-values satisfying
$\lambda+a-\frac{\delta}{2R}=\frac{j}{R}$
($j=0,1,2\ldots$)
are not eigenvalues, since they are canceled by simple poles of the cotangent function in the curly brackets. Therefore the eigenvalues are determined from the transcendental equation
\begin{equation}
    \lambda R=\pi |mR|^2 \cot[\pi R(\lambda +a) - \frac{\pi\delta}{2}].
    \label{spectrum}
\end{equation}
They define the finite masses of the 4-dimensional fermion eigenstates as functions of $m=N g \sigma$ and $A_5$,
if $\sigma_5$ and/or $A_5$ acquire non-vanishing vacuum expectation values.

\section{The effective potential and gap equation}

The generating functional of the system is given by
\begin{equation}
    Z= \int [\mathcal{D} \bar{\Psi}][\mathcal{D} \Psi][\mathcal{D}
    \sigma][\mathcal{D} \sigma^{*}] \; e^{i\int d^{4}x
    \mathcal{L}^{(4)}}.
\end{equation}
Next we perform the path-integration
over the fermion fields $\Psi$,
\begin{equation}
Z =\int [\mathcal{D} \sigma][\mathcal{D} \sigma^{*}] \; e^{-i N_f\int d^{4}x
    V_{\mathrm{eff}}(\sigma)}.
\end{equation}
As a result,  we obtain in the mean field approximation
(saddle point approximation in the leading order of the
$1/N_f$ expansion)
the effective potential
\begin{equation}
V_{\mathrm{eff}}(\sigma)= |\sigma|^2 - \int\frac{d^4
    k}{i(2\pi)^4}\ln \det(M+I\!\not\!k).
\end{equation}
Performing the Wick rotation we can rewrite the effective potential
as an integral over the Euclidean momentum
\begin{equation}
    V_{\mathrm{eff}}(\sigma)= |\sigma|^2 - \frac1{8\pi^2}\int_{0}^{\Lambda^2}
    d(k_E^2)\, k_E^2\ln \det(M^2+I k_E^2),
\end{equation}
where $\Lambda$ is the cutoff parameter.

To calculate the effective potential we used the following formula
\begin{equation}
    \det(M^2+I k_E^2)=\det(M+i I k_E)\det(M- i I k_E)
\end{equation}
and the explicit expressions for the determinants from formula
(\ref{det}).
After some algebra we obtain:
\begin{eqnarray}
    V_{\mathrm{eff}}(\sigma, a) &=& |\sigma|^2 - \frac1{4\pi^2}\int_{0}^{\Lambda}
    dx \, x^3 \Biggl\{ \ln \Bigl( x^2\left[\cosh 2\pi R x -\cos (2\pi R a-\pi\delta)\right] + 2\pi R x |m|^2  \sinh 2\pi R
    x
\nonumber \\
     &+&(\pi R |m|^2)^2 [\cosh 2\pi R x +\cos (2\pi R a -\pi\delta)] \Bigr)
     \Biggr\}.
     \label{eff}
\end{eqnarray}
The condensate $|\sigma|$ or the dynamical
mass quantity $|m|=Ng|\sigma|$ is
determined from the gap equation
\begin{equation}
{\partial V_{\mathrm{eff}}(\sigma, a)\over \partial \sigma}\,\,=\,\,0,
\end{equation}
or
\begin{eqnarray}
    &\displaystyle 1-\frac{g^2}{4\pi^2} \int_{0}^{\Lambda} dx \,x^3 \times \\
 &\displaystyle\frac{x\sinh 2\pi R x +\pi R |m|^2 [\cosh 2\pi R x + \cos (2\pi R a-\pi\delta) ]}{x^2[\cosh 2\pi R x -\cos (2\pi R a-\pi\delta)] + 2\pi R x |m|^2 \sinh 2\pi R
    x +(\pi R |m|^2)^2 [\cosh 2\pi R x +\cos(2\pi R a-\pi\delta)]} =0.
\end{eqnarray}

A non-trivial solution exists only if the coupling constant exceeds its critical value $g_c$, which is obtained from the gap equation by setting $|\sigma|=0$:
\begin{equation}
    1=\frac{g_c^2}{4\pi^2}\int_{0}^{\Lambda}dx \, x^2 \frac{\sinh 2\pi R x}{\cosh 2\pi R x -\cos (2\pi R a-\pi\delta)}.
\end{equation}
It is easy to obtain from this equation the asymptotic value
$g_c^{\ast}$ of the critical coupling at $R\to\infty$:
\begin{equation}
\label{cc}
 1=\frac{g_c^{\ast2} \Lambda^3}{12\pi^2},
\end{equation}
which evidently does not depend on the gauge potential $A_5$
nor on the phase factor $\delta$.

\begin{figure}[h]
$
    \begin{array}{cc}
        \subfigure[]{\includegraphics[width=80mm]{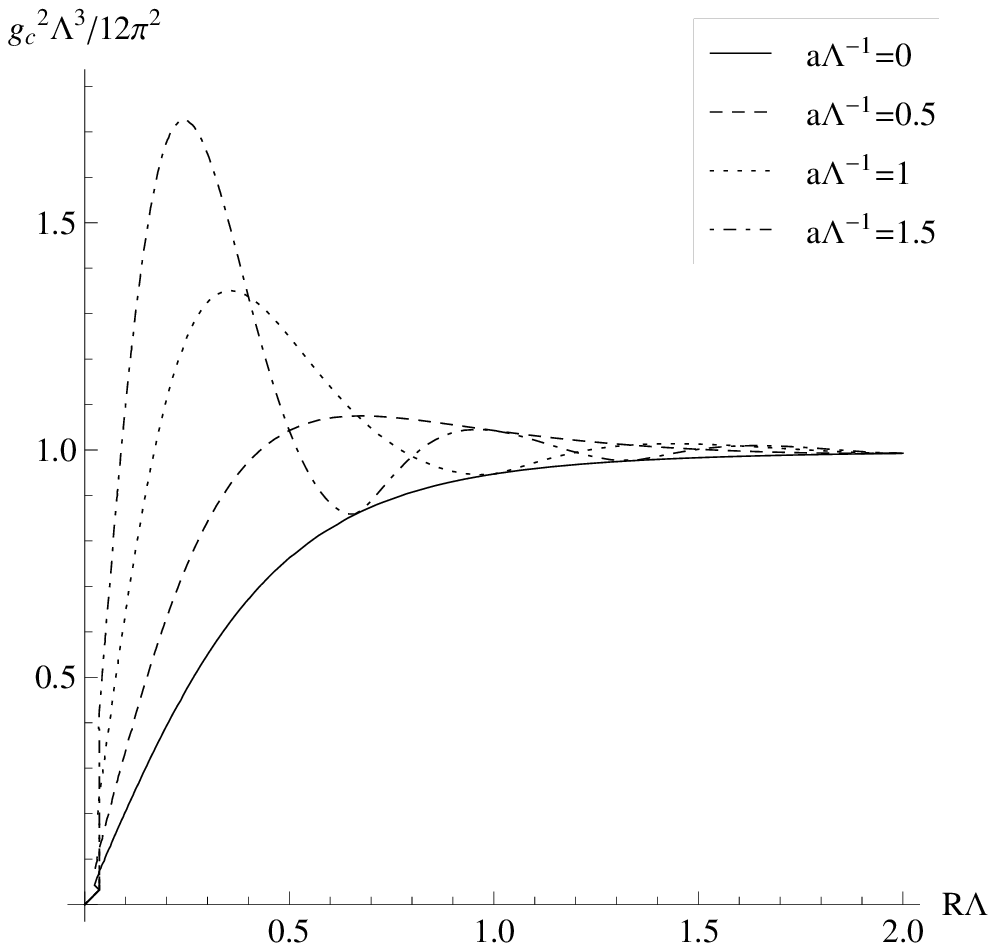}} &
        \subfigure[]{\includegraphics[width=80mm]{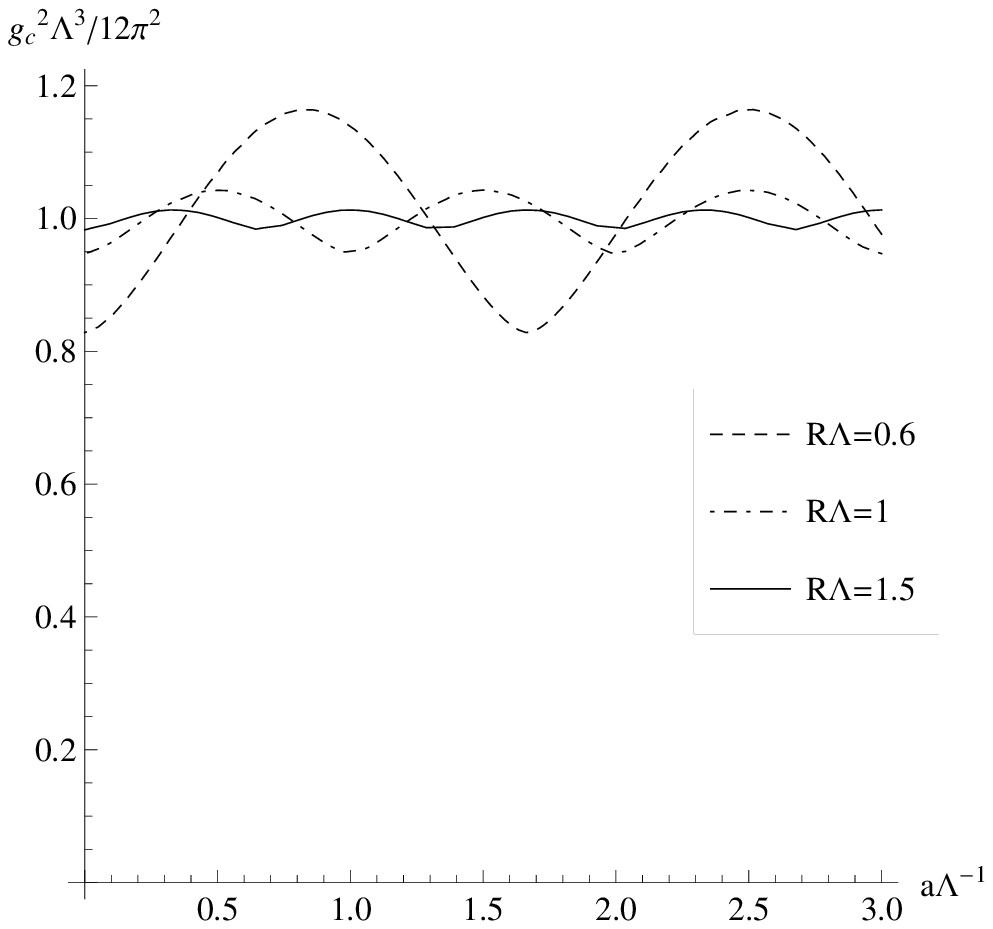}}
    \end{array}
$
\caption{The critical coupling constant as function of $R\Lambda$
at different fixed values $a\Lambda^{-1}$ (a) and as function of $a\Lambda^{-1}$ at different fixed values $R\Lambda$ (b) for the periodic boundary condition $\delta=0$.}
 \label{coupling}
\end{figure}

Fig.~\ref{coupling} shows the critical coupling constant as a function of the dimensionless radius $R\Lambda$ at different fixed
values $a\Lambda^{-1}$ and as a function
of the dimensionless potential $a\Lambda^{-1}$ at different fixed
values $R\Lambda$ for the periodic boundary condition, i.e. $\delta=0$.

It is seen from the Fig.~\ref{coupling}a that at small
radius ($R\Lambda<1$) the critical coupling oscillates, whereas at
large radius it tends to the asymptotic value (\ref{cc})
independent of the parameter $a$. In addition, the critical coupling
oscillates as a function of the potential $A_5$ (see the Fig.~\ref{coupling}b), however the amplitude of these oscillations
becomes small for large radius.

It is also clear that if the coupling constant only slightly exceeds
its critical value, the dynamical mass quantity $m$ is small and thus we can obtain light fermions in the considered model.

\begin{figure}[h]
$
    \begin{array}{cc}
        \subfigure[]{\includegraphics[width=80mm]{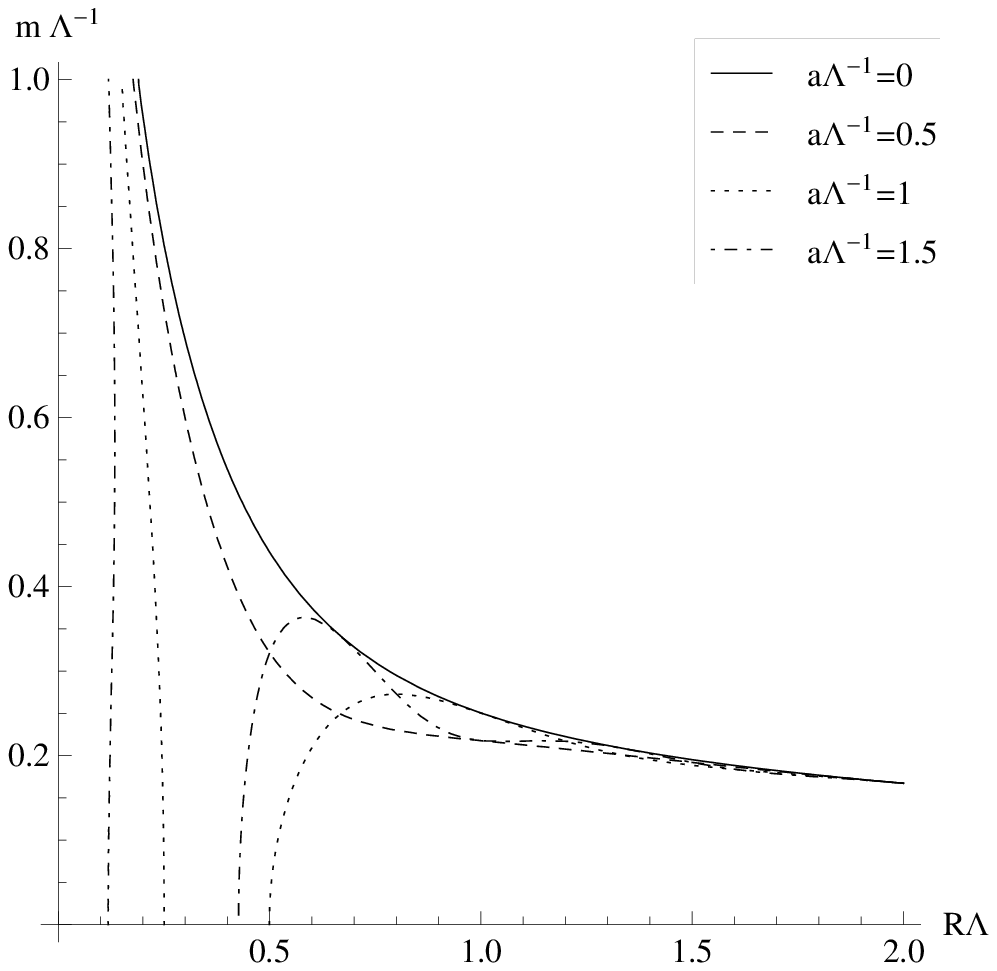}} &
        \subfigure[]{\includegraphics[width=80mm]{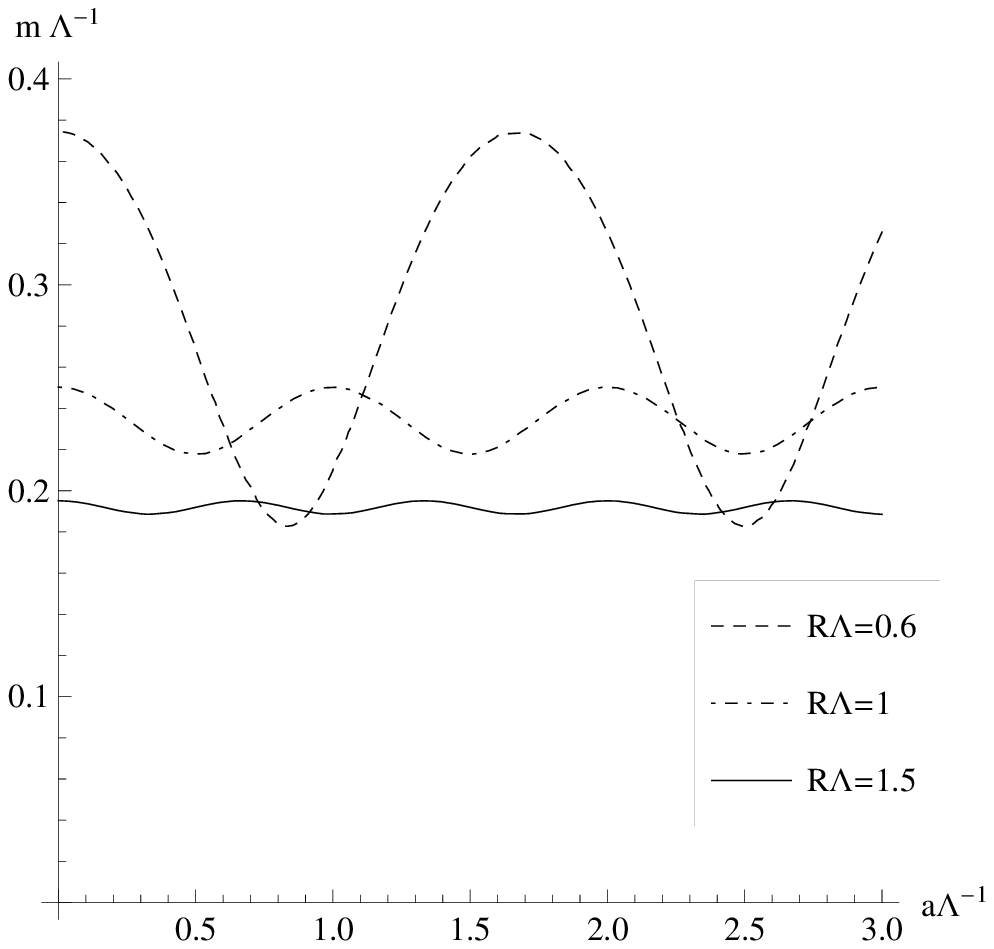}}
    \end{array}
$
\caption{The fermion mass quantity $m$ as function of $R\Lambda$
at different fixed values $a\Lambda^{-1}$ (a) and
as function of $a\Lambda^{-1}$ at different fixed values
$R\Lambda$ (b) for $g^2\Lambda^3/12\pi^2=1.25$ and $\delta=0$.}
 \label{m}
\end{figure}

The solution of the gap equation for the
mass quantity $m$ as function of the dimensionless compactification radius $R\Lambda$ for different fixed values of $a\Lambda^{-1}$ and as function of the dimensionless gauge field $a\Lambda^{-1}$ for different fixed values of $R\Lambda$ is depicted in Fig.~\ref{m} for
$g^2\Lambda^3/12\pi^2=1.25$ and $\delta=0$.
It is seen that at large radius $R$ ($R\Lambda>1$) the dynamical
quantity $m$ approaches its value corresponding to
a
vanishing gauge potential, $a=0$. Moreover, $m$ is an oscillating function of
the gauge potential $a$.

There seem to exist two solutions for
the mass quantity $m$ for $a\Lambda^{-1}=1$ or 1.5 as shown in Fig.~\ref{m}. The physical reason of this is the oscillation of the critical coupling on Fig.~\ref{coupling}a for $R\Lambda<1$. The curves for $a\Lambda^{-1}=1$ or 1.5 consist of two pieces, because at some values of $R\Lambda$ the critical coupling exceeds the fixed value $g^2\Lambda^3/12\pi^2=1.25$ (see Fig.\ref{coupling}a) and therefore there is no solution of the gap equation for these values of $R\Lambda$.

\section{Dynamical A-field}

Let us consider the gauge field 
component
$a$ as a dynamical variable
\cite{hosotani}. Then the extremum of the effective potential is
determined by the stationarity condition
\begin{eqnarray}
    &\displaystyle\frac{\partial V_{\mathrm{eff}}}{\partial a}=- \frac{R}{2\pi}\int_{0}^{\Lambda}
    dx \, x^3 \times \nonumber\\
& \times\displaystyle \frac{(x^2-(\pi R |m|^2)^2 \sin (2\pi R
a-\pi\delta)}{x^2[\cosh 2\pi R x -\cos (2\pi R a-\pi\delta)] + 2\pi
R x |m|^2 \sinh 2\pi R
    x +(\pi R |m|^2)^2 [\cosh 2\pi R x +\cos (2\pi R a-\pi\delta)]}=0,
\end{eqnarray}
The extremum is evidently achieved at $2Ra-\delta=n$, where $n \in
\mathbb{Z}$. The cases of even $n$ ($n=2k$) and odd $n$ ($n=2k+1$)
should be considered separately, because they lead
to different expressions for the effective potential. Therefore we have two types of solutions: $a=\frac{k}{R}+\frac{\delta}{2R}$ and
$a=\frac{k}{R}+\frac{1+\delta}{2R}$, ($k=0,1,2,\ldots$).
Recalling the large gauge
transformation (\ref{large}) it is clear that these two sets of solutions are equivalent to $a=\frac{\delta}{2R}$ and $a=\frac{1+\delta}{2R}$. Let us consider in the following only the cases of periodic and antiperiodic boundary conditions, i.e. $\delta=0$ and $\delta=1$. In the case $\delta=0$ we simply have two phases with trivial and nontrivial field $a$: $a=0$ and $a=\frac{1}{2R}$. It is easy to see that in the case of antiperiodic boundary condition these two phases just interchange. Therefore, in what follows we consider only the case of periodic boundary conditions, $\delta=0$.

\begin{figure}[h]
\includegraphics[width=90mm]{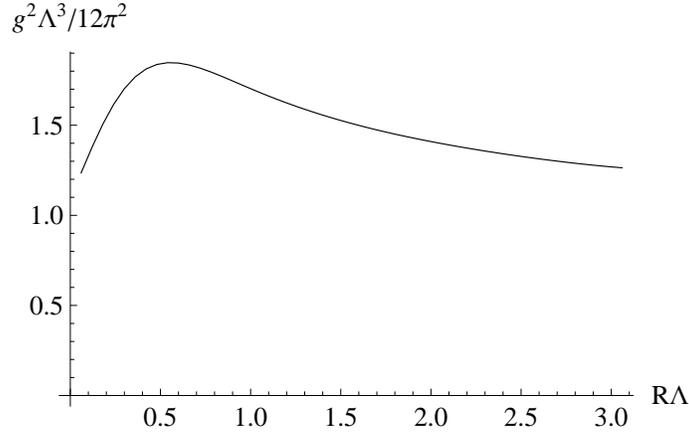}
\caption{The critical curve separates the phases with trivial and
non-trivial $a$. The region above (below) the curve corresponds to
$a=0$ ($a=\frac1{2R}$) for the periodic boundary
condition and vice versa for the antiperiodic one.}
 \label{curve}
\end{figure}

To obtain the true vacuum of the model one needs to compare the
minimum values
of the potentials $V_1(\sigma)$ and $V_2(\sigma)$
which follow from the effective potential (\ref{eff}) by letting
$a=\frac{\delta}{2R}$ and $a=\frac{1+\delta}{2R}$. The global minimum is realized at non-trivial values of $a$ depending on the coupling $g$ and radius $R$. The result of this comparison is presented in
Fig.~\ref{curve} where the critical curve which separates the phases
with $a=0$ and $a=\frac1{2R}$ is depicted. The region above (below) the curve corresponds to the phases with $a=0$ ($a=\frac1{2R}$) for
the periodic boundary condition and vice versa for the antiperiodic
one.

\begin{figure}[h]
$
\begin{array}{cc}
\subfigure[]{\includegraphics[width=8cm]{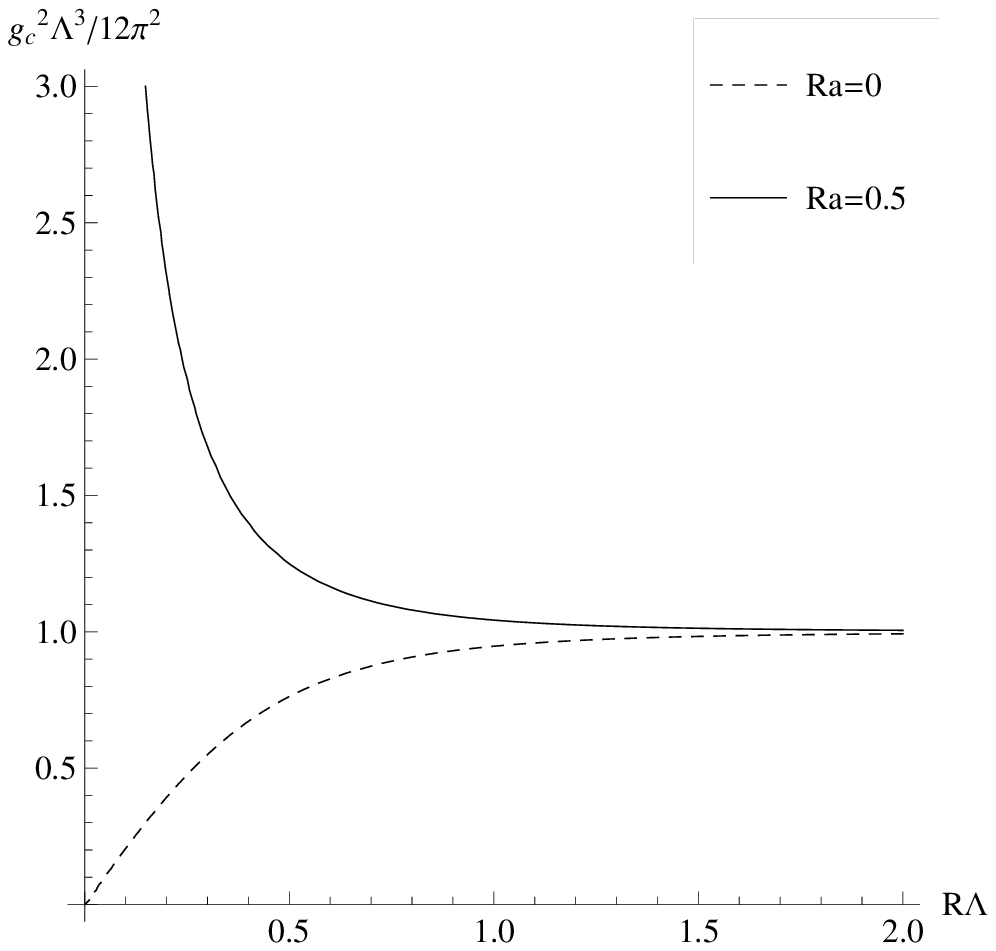}}&
\subfigure[]{\includegraphics[width=8cm]{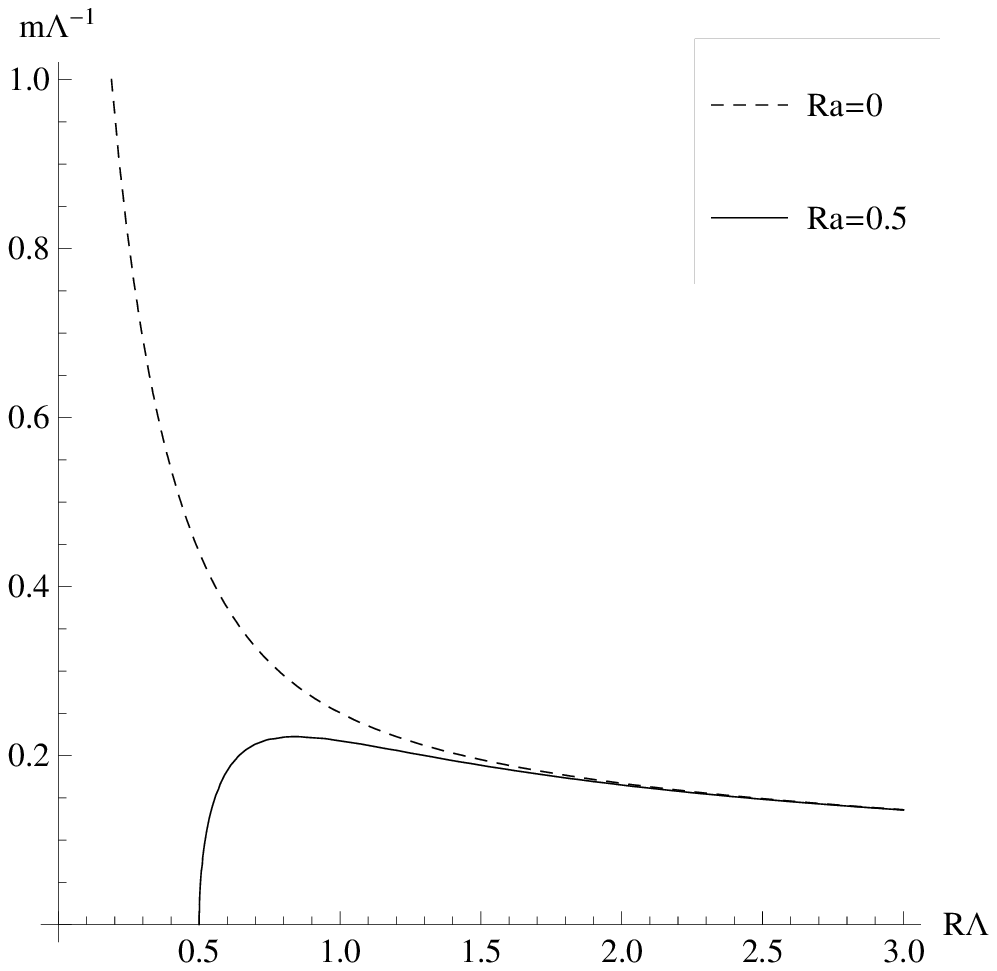}}
\end{array}
$
\caption{The critical coupling constant (a) and dynamical mass
  quantity $m$ as functions of the radius at
$g^2\Lambda^3/12\pi^2=1.25$ (b).}
 \label{cc1}
\end{figure}

Let us compare the dynamical mass generation for trivial
and non-trivial $a$. The critical couplings for both phases are
shown in Fig.~\ref{cc1}a.

In both cases the nontrivial solution for the dynamical mass
quantity may exist only for those values of the coupling constants that lie above the appropriate critical curve. It is seen that at $a=0$ the critical curve starts from zero at $R\to0$ and therefore even if the coupling is fixed in such a way that it is less than the asymptotic value (\ref{cc}) the dynamical mass can still be nonzero for small radius
($R\Lambda\lesssim1$) (if $g>g_c^{\ast}$ the fermion mass exists at
all radii of extra dimension). For $a=1/(2R)$ the critical coupling goes
to infinity at $R\to0$ and thus the gap equation has a
nonzero solution only if $g>g_c^{\ast}$ and only for
a sufficiently large radius.
This situation is illustrated in Fig.~\ref{cc1}b where the typical
behavior of the dynamical mass quantity is represented as a function of the radius inside both phases at fixed coupling
$g^2\Lambda^3/12\pi^2=1.25$.

Let us recall that the mass spectrum of 4-dimensional fermions is determined from
(\ref{spectrum}). For $a=0$ we simply obtain
\begin{equation}
    \lambda R=\pi |mR|^2 \cot\pi R\lambda.
\end{equation}
We are interested here in a light fermion with mass
$\lambda\ll 1/R$. It
can be easily shown that the lowest Kaluza-Klein mode in this case
has the mass $\lambda=\pm m$. The dynamical mass $|m|$ is small
($|m|\ll 1/R$), if the
coupling is close to its critical value.

When $a=(2R)^{-1}$ we have for the spectrum
\begin{equation}
  \lambda R=\pi |mR|^2 \cot\pi( R\lambda+{1\over 2}).
\label{pm}
\end{equation}
For $|\pi( R\lambda+{1\over 2})|\ll 1$
we have
\[
\lambda R=-{1\over 4}\,-\,\sqrt{{1\over 16}+|mR|^2}\approx -{1\over 2}+ O(|mR|^2)\,\,\, {\rm for}\,\,\, |mR|^2\ll 1.
\]
Therefore in this case we have $\lambda \approx -{1\over 2R}=-a$. Since the cotangent is periodical with the period $\pi$, we may rewrite (\ref{pm}) in the form
 \begin{equation}
  \lambda R=\pi |mR|^2 \cot\pi( R\lambda-{1\over 2}),
\label{pm1}
\end{equation}
with the corresponding approximate solution $\lambda \approx +{1\over  2R}=a$.
It is seen that independently of  
any nonzero value of $|m|>0$
the lowest
fermion
mode is
now
massless: $\lambda=0$. The higher modes have the masses $\lambda
\approx \pm {1\over 2R}=\pm a$,   and other Kaluza-Klein modes have even higher masses
$\lambda_{n}\approx \frac n R$.

\section{Summary}
In this paper
we have studied the dynamical mass generation
of fermions
in
a 5D model
with one compact extra dimension and
with bulk fermions in interaction with fermions
living
on a 3-brane
under the influence of
a constant 
gauge field component
$A_5$.
If $A_5$ is considered as a background
parameter,
then for sufficiently large radii the dynamical mass becomes
independent of
the
gauge field and approaches its value corresponding to
a vanishing gauge potential. Moreover, the dynamical mass is an
oscillating function of
the field 
component
$A_5$
with an amplitude of
oscillations
that decreases with growing radius $R$.

On the other hand,
if we consider the gauge potential as a dynamical variable,
then it may acquire trivial or nontrivial values depending on the
choice of the coupling constant $g$.
In particular,
it was demonstrated
that for a vanishing
gauge field
and a coupling constant close to its critical value,
it is possible to obtain
a light fermion mode whose mass is lower than the inverse radius,
which is
in correspondence with the results of \cite{abe}.
However, if $A_5$
acquires a nonvanishing value,
then the lowest
fermion mode
is massless independently of the
dynamical mass $m$.
Other possible modes have masses $\lambda\approx \pm{1\over 2R}=\pm a$, and
higher Kaluza-Klein modes have even higher masses $\lambda_{n}\approx
\frac n R$.  
The investigated interplay of the influence of Kaluza-Klein fermions and bulk
gauge fields on the generation of dynamical masses of fermions
living on a brane seems to us conceptually interesting and worth to be
generalized to other, phenomenologically more realistic models.

\section*{Acknowledgments}
V.Ch.Zh. and A.V.T. would like to thank
M.Mueller-Preussker and the members of the particle theory
group at HU-Berlin for their hospitality.
They  are grateful to
{\it Deutscher Akademischer Austauschdienst}
(DAAD) for financial support.

\end{document}